\documentclass[twocolumn,prl,amsmath,amssymb,showpacs,superscriptaddress,floatfix]{revtex4}
\usepackage{graphicx}% Include figure files
\usepackage{bm}% bold math

\usepackage{epsf}
\begin{document}
%\draft
\title{Can random pinning change the melting scenario of two-dimensional core-softened potential system?}%:
%First-Order versus Continuous Transition }

\author{E. N. Tsiok}
\affiliation{ Institute for High Pressure Physics RAS, 142190
Kaluzhskoe shosse, 14, Troitsk, Moscow, Russia}

\author{D.E. Dudalov}
\affiliation{ Institute for High Pressure Physics RAS, 142190
Kaluzhskoe shosse, 14, Troitsk, Moscow, Russia}

\author{Yu. D. Fomin}
\affiliation{ Institute for High Pressure Physics RAS, 142190
Kaluzhskoe shosse, 14, Troitsk, Moscow, Russia}
\affiliation{Moscow Institute of Physics and Technology, 141700
Moscow, Russia}

\author{V. N. Ryzhov}
\affiliation{ Institute for High Pressure Physics RAS, 142190
Kaluzhskoe shosse, 14, Troitsk, Moscow, Russia}
\affiliation{Moscow Institute of Physics and Technology, 141700
Moscow, Russia}

\date{\today}

\begin{abstract}
In experiments the two-dimensional systems are realized mainly on
solid substrates which introduce quenched disorder due to some
inherent defects. The defects of substrates influence the melting
scenario of the systems and have to be taken into account in the
interpretation of the experimental results. We present the results
of the molecular dynamics simulations of the two dimensional
system with the core-softened potential in which a small fraction
of the particles is pinned, inducing quenched disorder.The
potentials of this type are widely used for the qualitative
description of the systems with the water-like anomalies. In our
previous publications it was shown that the system demonstrates an
anomalous melting scenario: at low densities the system melts
through two continuous transition in accordance with the
Kosterlitz-Thouless-Halperin-Nelson-Young (KTHNY) theory with the
intermediate hexatic phase, while at high densities the
conventional first order melting transition takes place. We find
that the well-known disorder-induced widening of the hexatic phase
occurs at low densities, while at high density part of the phase
diagram random pinning transforms the first-order melting into two
transitions: the continuous KTHNY-like solid-hexatic transition
and first-order hexatic-isotropic liquid transition.

\end{abstract}


\maketitle

Despite almost forty years of investigations, the controversy
about the microscopic nature of melting in two dimensions ($2D$)
still lasts. In a crystal in contrast to an isotropic liquid two
symmetries are broken: translational and rotational. These two
symmetries are not independent, since a rotation of one part of an
ideal crystal with respect to another part disrupts not only the
orientational order but also the translational order. However, it
is possible to imagine the state of matter with orientational
order, but without the translational one. As it was shown by
Mermin \cite{mermin} in two dimensions the long-range
translational order can not exist because of the thermal
fluctuations and transforms to the quasi-long-range one. On the
other hand, the real long range orientational order does exist in
this case.

These ideas were used in the widely accepted KTHNY theory
\cite{kosthoul73,halpnel1,halpnel2,halpnel3} where it was proposed
that, in contrast to the $3D$ case where melting is the
first-order transition, the $2D$ melting can occur through two
continuous transitions. In the course of the first transition the
bound dislocation pairs dissociate at some temperature $T_m$
transforming the quasi-long range translational order into the
short-range one, and long-range orientational order into the
quasi-long range order. At this transition the decay of the
translational correlation function will change from algebraic to
exponential, and the orientational correlation function will obey
the algebraic decrease. The new intermediate phase with the
quasi-long range orientational order is called the hexatic phase.
After consequent dissociation of the disclination pairs at some
temperature $T_i$ the hexatic phase transforms into the isotropic
liquid with the exponential decay of the orientational
correlations.

The KTHNY theory seems universal and applicable to all system,
however, despite numerous experimental and simulation studies, it
could not be consistently verified or refuted. The KTNHY scenario
was unambiguously experimentally confirmed for superparamagnetic
colloidal particles interacting via long-range dipolar interaction
\cite{keim1,zanh,keim2,keim3,keim4}. On the other hand, in $2D$
the first-order melting without hexatic phase is also possible
\cite{chui83,ryzhovJETP}. The possible mechanisms of the
first-order melting are the formation of the grain boundaries
\cite{chui83} and the dissociation of the disclination quadrupoles
\cite{ryzhovJETP}. It is now widely believed that the melting
process strongly depends on the pair interaction of the system.
Numerous experimental and simulation studies demonstrate that the
systems with very short range or hard core potentials melt through
weak first-order transition, while the melting scenarios for the
soft repulsive particles favor the KTHNY theory. However,
controversies still exist even for the same systems, like, for
example, for hard spheres
\cite{rto1,RT1,RT2,prest2,DF,prest1,strandburg92,binderPRB,mak,binder,LJ}.

Recently, a number of papers have appeared where another melting
scenario was proposed \cite{foh1,foh2,foh3,foh4,foh5,foh6,foh7}.
It was argued that the basic hard disk model demonstrates the
two-stage melting transition with a continuous solid-hexatic
transition but a first-order hexatic-liquid transition
\cite{foh1,foh2,foh3}. In Ref. \cite{foh5} it was shown that in
$2D$ Yukawa system there is a two-phase coexistence region between
the stable hexatic phase and isotropic liquid. Kapfer and Krauth
\cite{foh4} have analyzed the effect of the potential softness on
the melting scenarios. They have proposed that the soft sphere
system with the potential of the form $U(r)=(\sigma/r)^n$ melts in
accordance with the KTHNY theory for $n\leq 6$, while for $n>6$
the two-stage melting transition takes place with the continuous
solid-hexatic transition and the first-order hexatic-liquid one.

In experiment, planar confinement is typically realized by
adsorption on solid substrates which introduce frozen-in
(quenched) disorder due to some roughness or intrinsic defects.
Quenched disorder can influence the crystallization/melting
scenario of two-dimensional systems.  As it was shown in Refs.
\cite{nel_dis1,nel_dis2} (see also \cite{dis3,dis4}), the KTHNY
melting scenario persists in the presence of weak disorder. The
temperature of the hexatic-isotropic liquid transition $T_i$ is
almost unaffected by disorder, while the melting temperature $T_m$
drastically decreases with increasing disorder
\cite{nel_dis1,nel_dis2,dis3,dis4}, and the stability range of the
hexatic phase widens. Recently these predictions have been tested
by experiment and simulations of the behavior of the
superparamagnetic colloidal particles \cite{keim3,keim4}.

In this paper, we present a computer simulation study of $2D$
phase diagram of the previously introduced core softened potential
system \cite{jcp2008,wepre,we_inv,we2011,RCR,we2013-2} in the
presence of the quenched disorder. The core-softened potentials
are widely used for the qualitative description of systems
demonstrating the water-like anomalous behavior, including
density, structural and diffusion anomalies, liquid-liquid phase
transitions, glass transitions, melting maxima
\cite{jcp2008,wepre,we_inv,we2011,RCR,we2013-2,buld2009,fr1,
barbosa,barbosa1,buld2d,scala}.

The system we study in the present simulations is described by the
potential
\cite{jcp2008,wepre,we_inv,we2011,RCR,we2013-2,dfrt1,dfrt2,dfrt3}:
\begin{equation}
U(r)=\varepsilon\left(\frac{\sigma}{r}\right)^{n}+\frac{1}{2}\varepsilon\left(1-
\tanh(k_1\{r-\sigma_1\})\right). \label{1}
\end{equation}
where $n = 14$ and $k_1\sigma = 10.0$. $\sigma$ is the hard-core
diameter. We simulate the systems with the soft-core diameter
$\sigma_1/\sigma = 1.35$. As it was shown in our previous
publications \cite{dfrt1,dfrt2,dfrt3,dfrt4}, the phase diagram of
the system consists of three different crystal phases (see also
similar picture in \cite{prest2}), one of them has square symmetry
and the other two are the low density and high density triangular
lattices (see, for example, Fig. 6 (a) in \cite{dfrt3}). Melting
of the low density triangular phase is a continuous two-stage
transition, with a narrow intermediate hexatic phase, in
accordance with the KTHNY scenario. At high density part of the
phase diagram, the square and triangular phases melt through one
first order transition. At higher temperatures and high density
there is one first order transition between trianglar solid and
isotropic liquid.
%The
%thermodynamic and dynamic anomalies do exist in this case,
%however, the order of this anomalies is inverted in comparison
%with the three-dimensional case \cite{dfrt3}.

%\begin{figure}

%\includegraphics[width=8cm]{Fig1.eps}%

%\caption{\label{fig:fig1} (Color online) Phase diagram of the
%system with the potential (\ref{1}) in the $\rho-T$ plane, where
%the triangular (T) and square (S) phases are shown. In the
%low-density part of the phase diagram the lines of solid-hexatic
%(open squares) and hexatic-liquid (filled squares) transitions are
%shown.}
%\end{figure}

In the remainder of this paper we use the dimensionless
quantities, which in $2D$ have the form: $\tilde{{\bf r}}\equiv
{\bf r}/\sigma$, $\tilde{P}\equiv P \sigma^{2}/\varepsilon ,$
$\tilde{V}\equiv V/N \sigma^{2}\equiv 1/\tilde{\rho}, \tilde{T}
\equiv k_BT/\varepsilon, \tilde{\sigma_1}=\sigma_1/\sigma$. In the
rest of the article the tildes will be omitted.

We use the molecular dynamics simulations of the system in $NVT$
and $NVE$ ensembles (LAMMPS package \cite{lammps}) with the number
of particles equal to $20000$. Quenched disorder is introduced by
pinning a randomly chosen subset of particles to the random
positions and let them to be immobile for the entire simulation
run. We make the simulations of 10 independent replicas of the
system with different distributions of random pinned pattern and
after that average the thermodynamic functions over replicas. We
calculate the pressure $P$ versus density $\rho$ along the
isotherms, and the correlation functions $G_6(r)$ and $G_T(r)$ of
the bond orientational $\psi_6$ and translational $\psi_T$ order
parameters (OPs), which characterize the overall orientational and
translational order.

We define the translational order parameter $\psi_T$ (TOP), the
orientational order parameter $\psi_6$ (OOP), the
bond-orientational $G_6(r)$ (OCF) and translational $G_T(r)$ (TCF)
correlation functions in the conventional way
\cite{prest2,halpnel1,halpnel2,binder,binderPRB,LJ,prest1,foh7}.

TOP can be written in the form
\begin{equation}
\psi_T=\frac{1}{N}\left<\left<\left|\sum_i e^{i{\bf G
r}_i}\right|\right>\right>_{rp}, \label{psit}
\end{equation}
where ${\bf r}_i$ is the position vector of particle $i$ and {\bf
G} is the reciprocal-lattice vector of the first shell of the
crystal lattice. The translational correlation function can be
computed in accordance with the definition \cite{halpnel1,
halpnel2, foh7}:
\begin{equation}
G_T(r)=\left<\frac{<\exp(i{\bf G}({\bf r}_i-{\bf
r}_j))>}{g(r)}\right>_{rp}, \label{GT}
\end{equation}
where $r=|{\bf r}_i-{\bf r}_j|$ and $g(r)=<\delta({\bf
r}_i)\delta({\bf r}_j)>$  is the pair distribution function. The
second angular brackets $<...>_{rp}$ correspond to the averaging
over the random pinning. In the solid phase the long range
behavior of $G_T(r)$ has the form \cite{halpnel1, halpnel2}
$G_T(r)\propto r^{-\eta_T}$ with $\frac{1}{4}\leq \eta_T \leq
\frac{1}{3}$.

To identify the existence  the orientational order and the hexatic
phase, we define the local order parameter, which measures the
$6$-fold orientational ordering, in the following way:
\begin{equation}
\Psi_6({\bf r_i})=\frac{1}{n(i)}\sum_{j=1}^{n(i)} e^{i
n\theta_{ij}}\label{psi6loc},
\end{equation}
where $\theta_{ij}$ is the angle of the vector between particles
$i$ and $j$ with respect to a reference axis and the sum over $j$
is counting $n(i)$ nearest-neighbors of $j$, found from the
Voronoi construction. The global OOP is obtained as an average
over all particles and random pinning:
\begin{equation}
\psi_6=\frac{1}{N}\left<\left<\left|\sum_i \Psi_6({\bf
r}_i)\right|\right>\right>_{rp}.\label{psi6}
\end{equation}

The orientational correlation function $G_6(r)$ (OCF) is given by
the equation analogous to Eq. \ref{GT}:
\begin{equation}
G_6(r)=\left<\frac{\left<\Psi_6({\bf r})\Psi_6^*({\bf
0})\right>}{g(r)}\right>_{rp}, \label{g6}
\end{equation}
where $\Psi_6({\bf r})$ is the local bond-orientational order
parameter (\ref{psi6loc}). In the hexatic phase there is a
quasi-long range order with the algebraic decay of the
orientational correlation function $G_6(r) \propto r^{-\eta_6}$
with $0\leq \eta_6 \leq \frac{1}{4}$
\cite{halpnel1,halpnel2,halpnel3}.

\begin{figure}

\includegraphics[width=8cm]{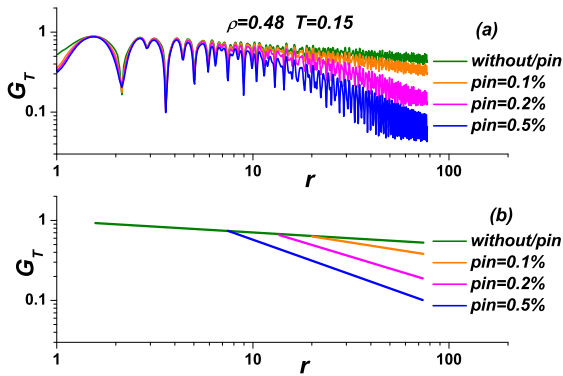}%

\caption{\label{fig:fig2} (Color online) (From top to bottom)
Translational correlation functions (a) and enveloping curves (b)
for: 1). no random pinning; 2). concentration of pinning centers
equal to $0.1\%$; 3). $0.2\%$; 4). $0.5\%$ ($\rho=0.48, T=0.15$).
}
\end{figure}

Before to proceed with the study of the phase diagram of system
(\ref{1}) in the presence of the random pinning, let us consider
the influence of disorder on the orientational and translational
correlation functions. In accordance with the results of Nelson
and coworkers \cite{nel_dis1, nel_dis2}, we did not find any
qualitative change in the behavior of $G_6(r)$ (Eq. (\ref{g6})) in
the presence of pinning. In contrast with $G_6(r)$, we found the
qualitatively different behavior of the translational correlation
function $G_T(r)$ (Eq. (\ref{GT})). In Fig.~\ref{fig:fig2} we
present $G_T(r)$ for $\rho=0.48, T=0.15$ without pinning and for 3
different values of the concentration of pinning centers: $0.1\%,
0.2\%, 0.5\%$ (Concentration $0.1\%$ corresponds to 20 particles
frozen-in in the random positions in the system). Density and
temperature $\rho=0.48, T=0.15$ correspond to the point on the
phase diagram deeply inside in the low density solid phase.
Without pinning we have the conventional power law decay of TCF.
In the case of pinning, the slope of the enveloping line increases
at some crossover value $r_0$. The region $r<r_0$ corresponds to
the local order unaffected by random impurities, while asymptotic
behavior of TCF for $r>r_0$ is determined by the random pinning
\cite{nel_dis1}. From Fig.~\ref{fig:fig2} (b) one can see that, in
accordance with the intuitive picture, $r_0$ decreases with
increase of impurities concentration along with the increase of
the slope of the enveloping line. Further we will consider the
equality $\eta_T=1/3$ for the long range asymptote of TCF as the
solid-hexatic transition criterium. The hexatic-liquid transition
occurs at $\eta_6=1/4$.

\begin{figure}

\includegraphics[width=8cm]{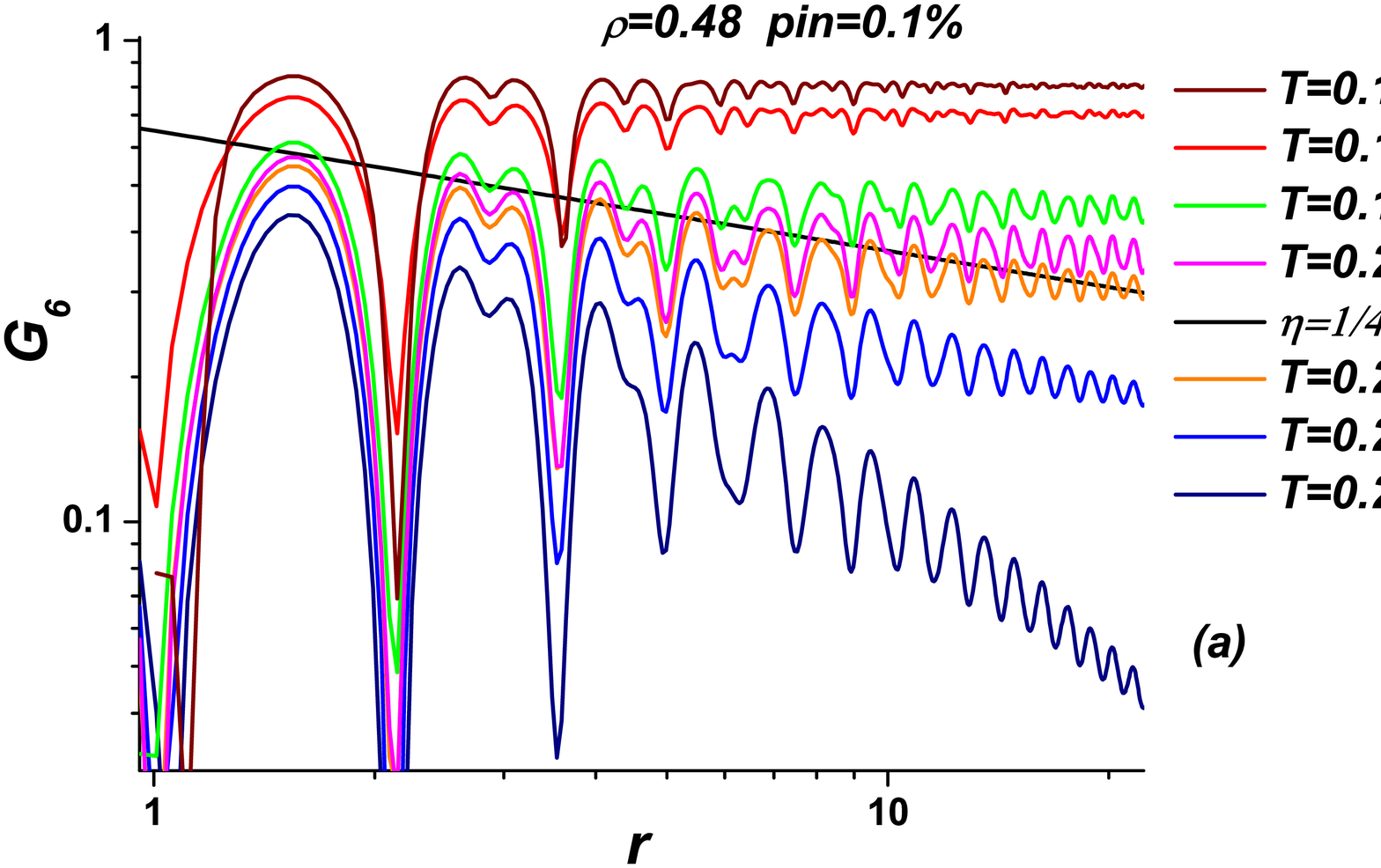}%

\includegraphics[width=8cm]{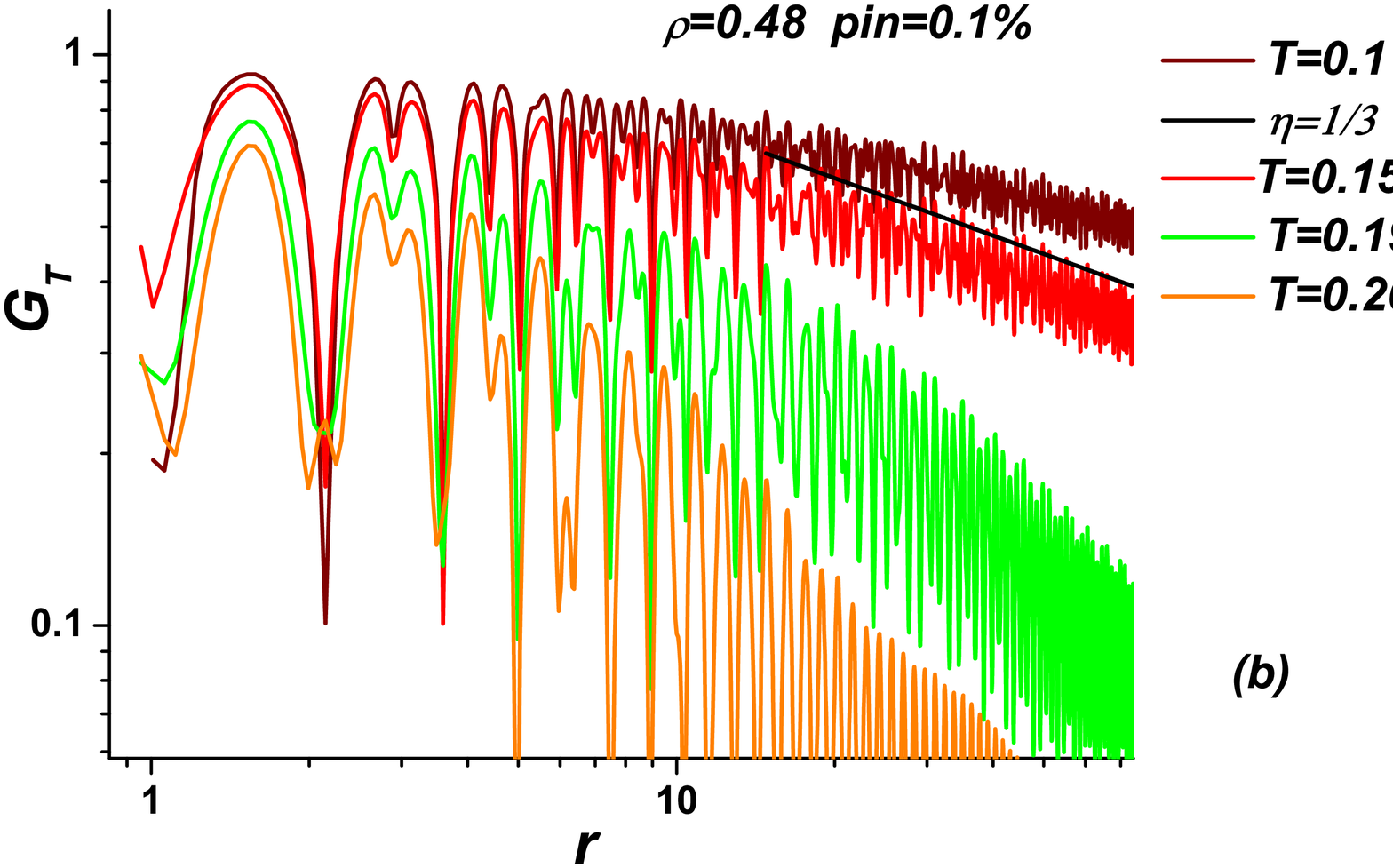}%

\caption{\label{fig:fig3} (Color online) OCF and TCF for different
temperatures at $\rho=0.48$ and concentration of the pinning
centers $0.1\%$}
\end{figure}

\begin{figure}%

\includegraphics[width=8cm]{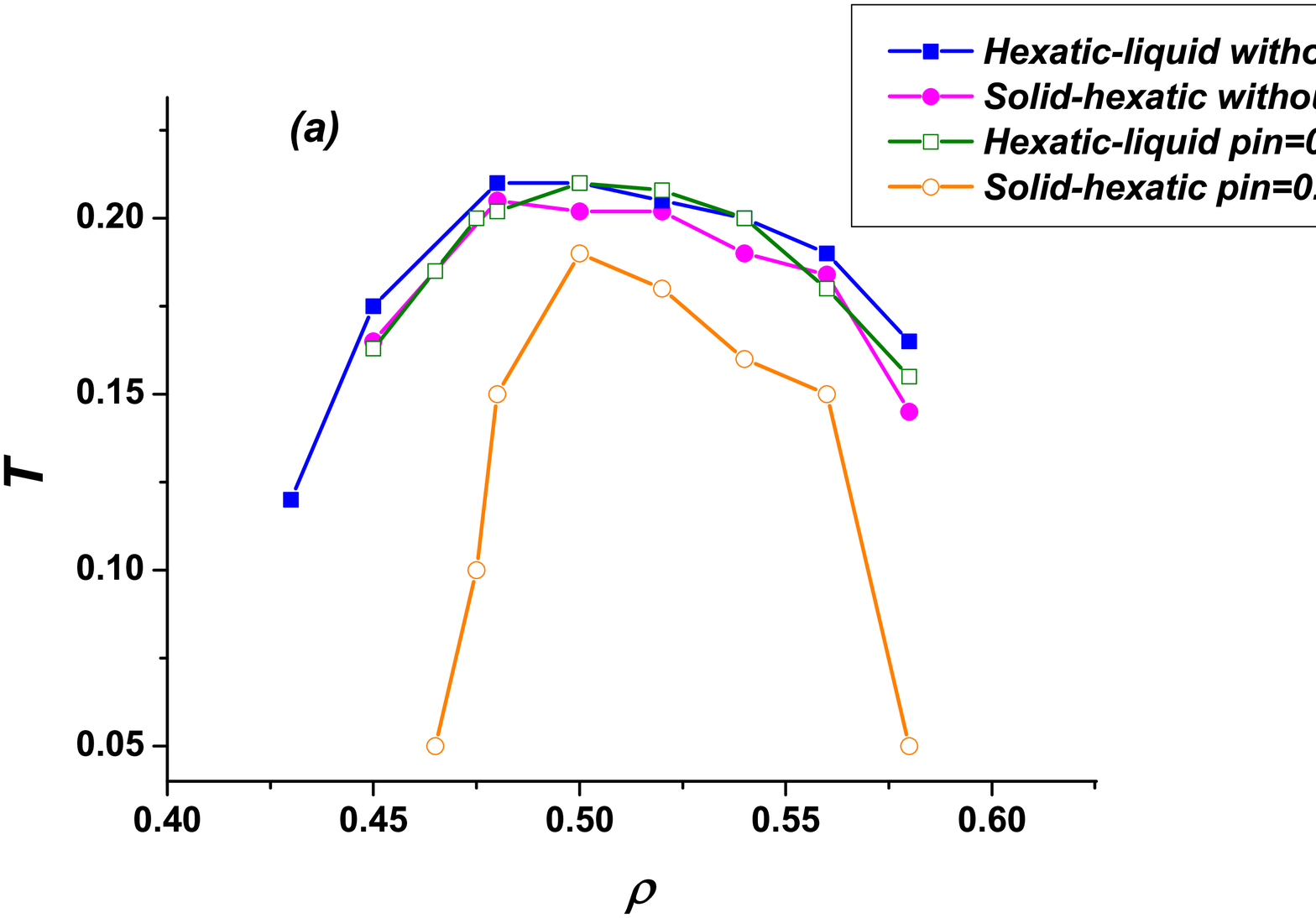}%

\includegraphics[width=8cm]{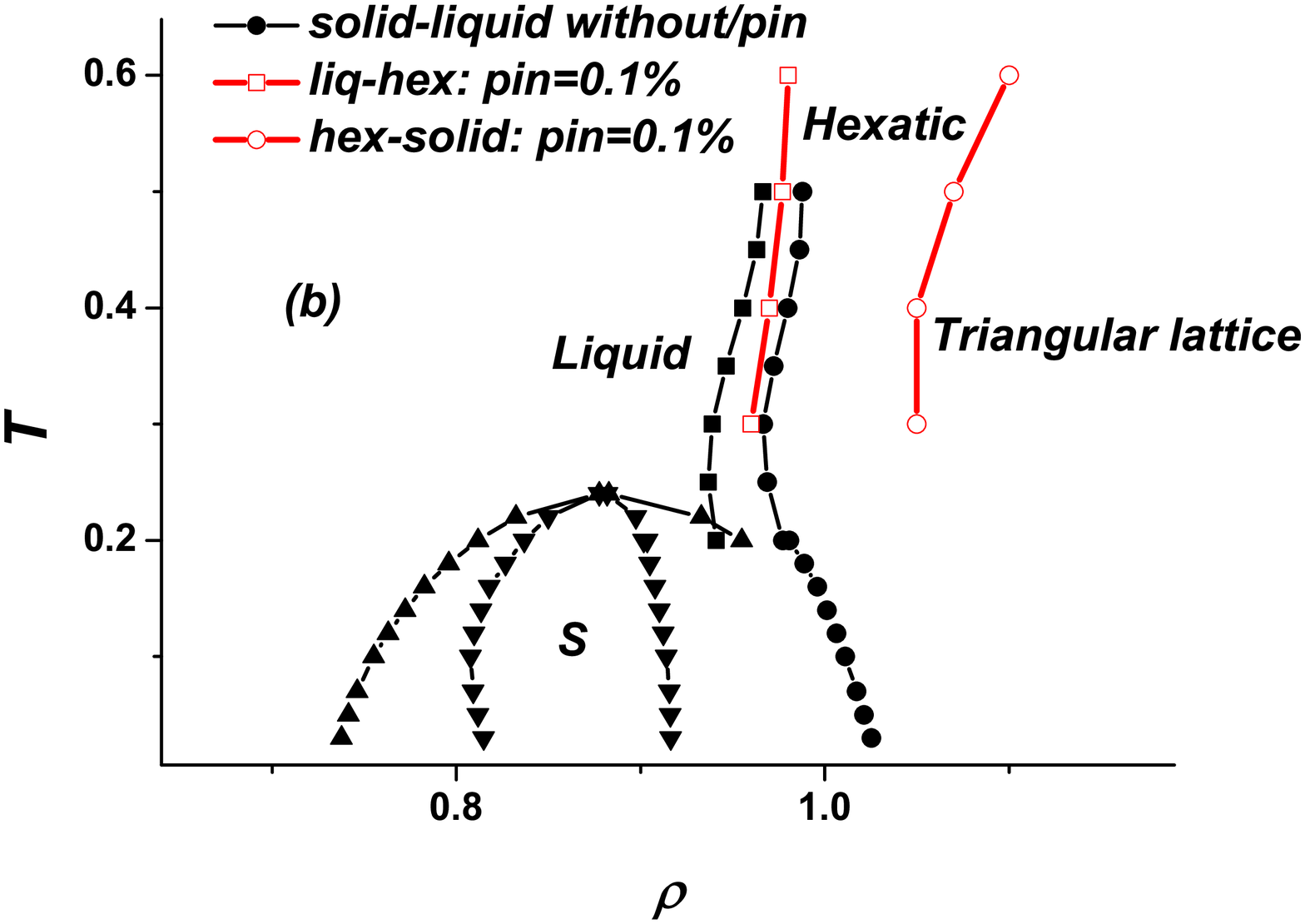}%

\caption{\label{fig:fig4} (Color online) Low density (a) and high
density (b) phase diagram of the system with the potential
(\ref{1}) without pinning (filled symbols) and with the
concentration of the pinning centers equal to $0.1\%$ (open
symbols).}
\end{figure}

Let us first consider the effect of pinning on the low density
part of the phase diagram where melting occurs in accordance with
the KTHNY theory. Fig.~\ref{fig:fig3} illustrates the behavior of
OCF and TCF for different temperatures at $\rho=0.48$ and the
concentration of the pinning centers equal to $0.1\%$. Using
equations $\eta_T=1/3$ and $\eta_6=1/4$, we calculated the
low-density phase diagram in the presence of random pinning. The
phase diagram is shown in Fig.~\ref{fig:fig4} (a). One can see
that random pinning leaves almost unaffected the line of the
liquid-hexatic transition, while considerably changes the location
of hexatic-solid transition. As a result, in the presence of
random pinning the hexatic phase drastically widens in accordance
with the theoretical predictions \cite{nel_dis1,nel_dis2} (see
also \cite{keim3,keim4}).

%\begin{figure}%

%\includegraphics[width=8cm]{Fig5.eps}%

%\caption{\label{fig:fig5} (Color online) High density phase
%diagram of the system with the potential (\ref{1}) without pinning
%(filled symbols) and with the concentration of the pinning centers
%equal to $0.1\%$ (open symbols) for $T>0.3$.}
%\end{figure}

\begin{figure}%

\includegraphics[width=8cm]{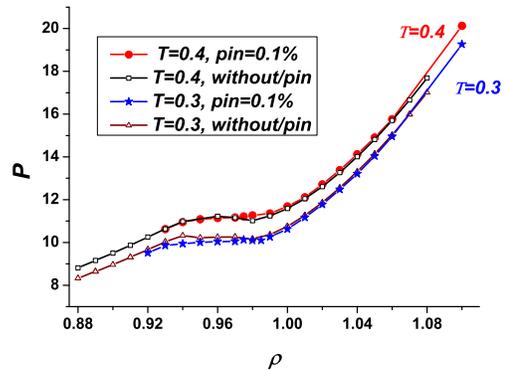}%

\caption{\label{fig:fig6} (Color online) Isotherms of the system
with the potential (\ref{1}) without pinning (open symbols) and
with the concentration of the pinning centers equal to $0.1\%$
(filled symbols) for $T=0.3$ and $T=0.4$.}
\end{figure}

The effect of random pinning on the first-order melting transition
is shown in Fig.~\ref{fig:fig4} (b). The temperatures higher than
$0.3$ are considered. In this temperature range the transition in
the pure system with the potential (\ref{1}) is of first order
\cite{dfrt1,dfrt2,dfrt3}. The lines determined from the equations
$\eta_T=1/3$ and $\eta_6=1/4$ are denoted by the open symbols. One
can see, that the line where $G_6(r)$ decays algebraically with
the exponent equal to $1/4$ is located inside the two-phase region
of the first-order melting transition of the system without
pinning. At the same time, in the presence of random pinning the
line of solid-hexatic transition obtained from condition
$\eta_T=1/3$ is shifted to higher densities. From this picture one
can conclude that random pinning transforms the single first-order
transition into two transition with rather wide hexatic phase. The
solid-hexatic transition is continuous, while the nature of the
transition from the hexatic phase to the isotropic liquid is
unclear. In order to elucidate this issue, we plot the isotherms
of the system in Fig.~\ref{fig:fig6} with and without random
pinning. One can see, that the van der Waals loops which are
characteristic of the first order transition are almost unchanged
by impurities. This means that the hexatic phase transforms into
isotropic liquid through first-order transition.

In conclusion, in this paper we present the computer simulation
study of melting transition in $2D$ core softened system in the
presence of random pinning. It is shown that at low densities,
where the system without pinning melts through two continuous
transitions in accordance with the KTHNY scenario, random disorder
does not change the character of the transition but drastically
widens the range of the hexatic phase.  At the same time, at high
densities where the conventional first order transition takes
place without random pinning, disorder drastically change the
melting scenario. The single first-order transition transforms
into two transitions, one of them (solid-hexatic) is the
continuous KTHNY-like transition, while the hexatic to isotropic
liquid transition probably occurs as the first order transition in
accordance with the recently proposed scenarios
\cite{foh1,foh2,foh3,foh5,foh4}. The possible mechanism for this
transition is the spontaneous proliferation of grain boundaries
\cite{chui83,foh3,foh5}.

It should be noted, that the nature of the first-order
liquid-hexatic transition is still puzzling. As it was shown in
\cite{RT1,RT2,dfrt4}, using the Landau-type expansion one can see
that the solid-liquid transition can be of first order in the case
of small enough thermal fluctuations and continuous KT-like, when
the singular fluctuations of the phase of the order parameter
takes place. The choice between these two possibilities depends on
the form of the potential and the thermodynamic parameters. In
contrast, in the case of the liquid-hexatic transition one can
show that both Landau expansion and KT mechanism lead to the
continuous transition, however, simulations predict the first
order transition. The strict theory like the KTHNY one is
necessary for explaining this controversy. The results of this
study can be useful for the qualitative understanding the behavior
of water confined in the hydrophobic slit nanopores
\cite{nat1,nat2}.

\bigskip

The authors are grateful to S.M. Stishov and V.V. Brazhkin for
valuable discussions. "We thank the Russian Scientifc Center at
Kurchatov Institute and Joint Supercomputing Center of Russian
Academy of Science for computational facilities. The work was
supported by the Russian Foundation For Basic Research (Grant No
14-02-00451).

\end{document}